\documentclass[aps,twocolumn,superscriptaddress,prl]{revtex4-1}
\usepackage{amsmath}
\usepackage{amssymb}
\usepackage{graphicx}
\usepackage{color,soul}
\newcommand{\er}{Erd\H{o}s-R\'{e}nyi}

\newcommand{\av}[1]{\langle #1 \rangle}

\newcommand{\dodo}[2]{\frac{\partial #1}{\partial #2}}

\begin{document}

\title{Efficient network immunization under
	limited knowledge}


\author{Yangyang Liu}
\thanks{Y. Liu and H. Sanhedrai contributed equally to this work}
\affiliation{Department of Systems Science,  College of Liberal Arts and Sciences, National University of Defense Technology, Changsha, Hunan 410073, China}
\author{Hillel Sanhedrai}
\thanks{Y. Liu and H. Sanhedrai contributed equally to this work}
\affiliation{Department of Physics, Bar-Ilan University, Ramat Gan 5290002, Israel}
\author{GaoGao Dong}
\email{To whom correspondence may be addressed. Email: dfocus.gao@gmail.com or lsheks@gmail.com}
\affiliation{Faculty of Science, Jiangsu University, Zhenjiang, Jiangsu 212013, China}
\author{Louis M. Shekhtman}
\email{To whom correspondence may be addressed. Email: dfocus.gao@gmail.com or lsheks@gmail.com}
\affiliation{Department of Physics, Bar-Ilan University, Ramat Gan 5290002, Israel}
\affiliation{Networks Science Institute, Northeastern University, Boston, MA 02115}
\author{Fan Wang}
\affiliation{Department of Physics, Bar-Ilan University, Ramat Gan 5290002, Israel}
\author{Sergey V. Buldyrev}
\affiliation{Department of Physics, Yeshiva University, New York, New York 10033, USA}
\author{Shlomo Havlin}
\affiliation{Department of Physics, Bar-Ilan University, Ramat Gan 5290002, Israel}

\date{\today}

\begin{abstract}
Targeted immunization or attacks of  large-scale networks has attracted significant attention by the scientific community. However, in real-world scenarios, knowledge and observations of the network may be limited thereby precluding a full assessment of the optimal nodes to immunize (or remove) in order to avoid epidemic spreading such as that of current COVID-19 epidemic. Here, we study a novel immunization strategy where only $n$ nodes are observed at a time and the most central between these $n$ nodes is immunized (or attacked). This process is continued repeatedly until $1-p$ fraction of nodes are immunized (or attacked). We develop an analytical framework for this approach and determine the critical percolation threshold $p_c$ and the size of the giant component $P_{\infty}$ for networks with arbitrary degree distributions $P(k)$. In the limit of $n\to\infty$ we recover prior work on targeted attack, whereas for $n=1$ we recover the known case of random failure. Between these two extremes, we observe that as $n$ increases, $p_c$ increases quickly towards its optimal value under targeted immunization (attack) with complete information. In particular, we find a new scaling relationship between $|p_c(\infty)-p_c(n)|$ and $n$ as $|p_c(\infty)-p_c(n)|\sim n^{-1}\exp(-\alpha n)$. For Scale-free (SF) networks, where $P(k)\sim k^{-\gamma}, 2<\gamma<3$, we find that $p_c$ has a transition from zero to non-zero when $n$ increases from $n=1$ to order of $\log N$ ($N$ is the size of network). Thus, for SF networks, knowledge of order of $\log N$ nodes and immunizing them can reduce dramatically an epidemics. 
\end{abstract}
\maketitle

Networks play a crucial role in many diverse systems \cite{newman2003structure,albert2002statistical,cohen2010complex,cohen2003efficient,brockmann2013hidden,liu2012core,liu2011controllability,watts2002simple}. Connectivity of components is critical for maintaining the functioning of infrastructures like the internet \cite{albert1999internet} and transportation networks \cite{toroczkai2004network},  as well as for understanding immunization against epidemics \cite{pastor2001epidemic}  and the spread of information in social systems \cite{watts1998collective}.  Due to this importance, researchers have long focused on how a network can be optimally immunized or fragmented to prevent epidemics or to maintain infrastructure resilience \cite{gallos2005stability,huang2011robustness,neumayer2011assessing,helbing2013globally,cohen2010complex,eubank2004modelling}. Many approaches have used percolation theory from statistical physics to prevent the spread of virus or assess network resilience under the infection or failure of some fraction of nodes or links \cite{stauffer2014introduction,stanley1971phase,buldyrev2010catastrophic,newman2001random,gao2012networks,Dong_2013_052804,shekhtman2016recent,bunde2012fractals,dong2018resilience,coniglio1977percolation,morone2015influence}. 

Early studies in networks found that immunizing real networks against an epidemic is highly challenging due to the existence of hubs which prevent eradication of the virus even if many nodes are immunized \cite{albert2000error,cohen2000resilience,callaway2000network}. At the same time, if these hub nodes, largest degree nodes, are targeted, the network can reach immunity \cite{albert2000error,callaway2000network,cohen2001breakdown}.  However, these previous models of targeted immunization have assumed full knowledge of the network structure which in many cases is not available. Recent research has shown that even those in control of a network often have knowledge only of a small part of the whole network structure \cite{yan2015spectrum,PhysRevLett.109.258701,Liu2460}. This was particularly demonstrated with the current COVID-19 epidemic where the detailed social network of individuals is unknown. 

In this Letter, we study targeted immunization or removal in networks with limited knowledge. We assume that at each stage, a number $n$ nodes are observed and the node with highest degree is immunized and thus unable to continue spreading infection. This procedure is repeated until $1-p$ fraction of nodes are immunized. In particular, our model could apply to a situation where several cooperative teams are sent to immunize a network and each team has access to information on a small subset ($n$ nodes) of the network.  We develop a theoretical framework for this model of immunization with limited information using percolation theory for networks with arbitrary degree distribution. In the limit of
$n = N$ we recover prior work on targeted attack \cite{cohen2001breakdown}, whereas for $n = 1$ we recover the case of random failure \cite{callaway2000network,cohen2000resilience}. We observe excellent agreement between our theoretical and simulation results for the $n$ dependence of the critical threshold $p_c$ and the size of the giant component $P_\infty$ for $p>p_c$. The giant component and $p_c$ characterize the efficiency of the immunization. The smaller is the giant component, the immunization   strategy is better. The larger is $p_c$  the immunization is more efficient since less immunization doses are needed to stop the epidemics. We find an analytical relationship between $n$ and $p_{c}$ for both \er \ (ER) and Scale-Free (SF) networks. Surprisingly, we also find that $p_c$ quickly reaches a plateau even for relatively small $n$, after which increasing $n$ has negligible effect on $p_c$. This means that immunization with small $n$ (not knowing the whole structure) can dramatically improve the immunization. 

Let $G(V,E)$  be a network where $V$ and $E$ are the set of nodes and edges, respectively. $N=|V|$ is the number of nodes in the network. We assume that a preventer or attacker has limited knowledge of the overall network structure and instead possesses only limited information on several nodes. Specifically, we randomly select $n$ nodes for which the preventer or attacker is assumed to have information on the node degree. The preventer or attacker then targets the node with the highest degree among these $n$. This procedure is then repeated until a $1-p$ fraction of nodes are immunized or removed from the network. 

\begin{figure}
	\centering
	\includegraphics[width=0.9\linewidth]{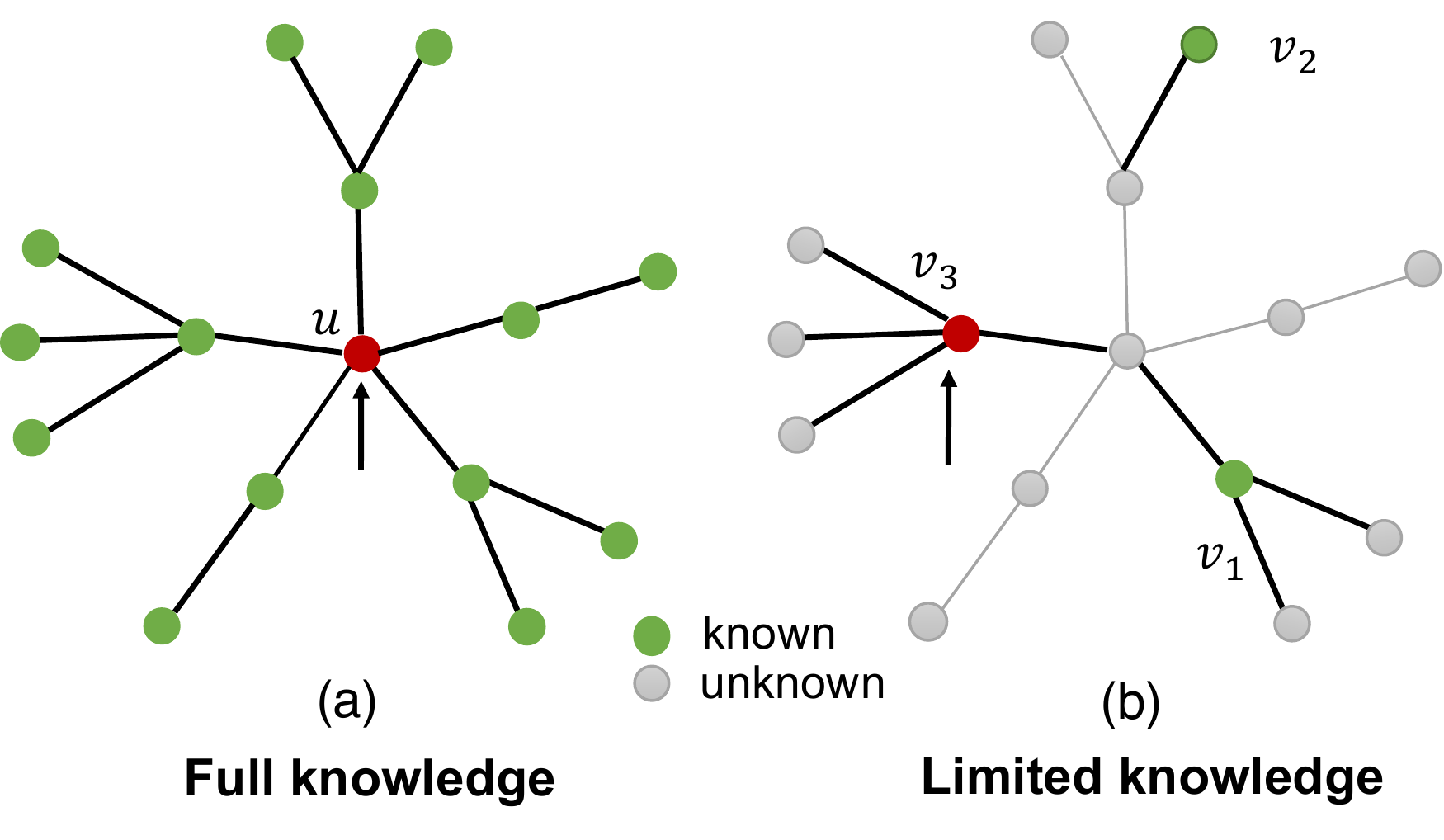}
	\caption{Schematic illustration of our limited knowledge immunization or attack strategy.  The preventer or attacker is able to observe the degree of nodes that are colored green, while the gray nodes are unknown.  (a) For the classical targeted immunization (attack), one has complete information on the global structure of the network and chooses the highest degree node ($u$) to immunize (remove). (b) Here the case of an individual with limited knowledge of the network is demonstrated. In this figure, we set $n=3$ and only degrees of nodes $v_1$, $v_2$ and $v_3$ are known. Given this limited information, the preventer would choose to immunize $v_3$, being unaware that an unobserved higher degree node exists. At the next immunization or attack, only nodes which have not been immunized or removed yet will be observed.}\label{fig:schematic}
\end{figure}

In Fig.\ref{fig:schematic}, the limited information immunization (attack) is illustrated together with global targeted immunization on a network. Here a total of $n=3$ nodes are observed. In panel (a), an individual with global information about the network structure chooses the highest degree node $u$ to immunize (or remove). However, in panel (b), the individual only knows at a time the degree of $3$ nodes in the network, i.e. $v_{1},v_{2},v_{3}$. Consequently, node $v_{3}$ with the highest degree $k=4$ (marked in red)  is immunized (or removed). 

Suppose the degree distribution of a network is given by $P(k)$ and $F(k)=\sum_{s=0}^{k}P(s)$ being the cumulative probability that the degree of a randomly chosen node is less than or equal to $k$. Furthermore, at an arbitrary time $t$ during the iterative percolation process, assume the distribution of the original degree (including the immunized neighbors) of the remaining nodes is $P(k,t)$. Then, the degree distribution of the node which is immunized at time $t$ is given by
\begin{equation}
	P_{r}(k,t)=F(k,t)^{n}-F(k-1,t)^{n}\equiv \Delta \left[F(k,t)^n\right],
	\label{eq:pk of one removed node}
\end{equation}
where $F(k,t)$ is the cumulative distribution of $P(k,t)$. This formula can be recognized as being derived from \emph{order statistics} giving the \emph{maximum} of several independent random variables \cite{hoel1954introduction}. For $k=0$, Eq.~\eqref{eq:pk of one removed node} becomes $P_{r}(0,t)=F(0,t)^{n}$. Hence we define $F(k-1,t)=0$, and then Eq.~\eqref{eq:pk of one removed node} is valid for $k\geq0$. 

In a limited knowledge immunization or attack, each node's immunization changes the degree distribution of the  remaining nodes in the following way
\begin{equation} \label{eq:N(t) first}
N(k,t+1)=N(k,t)-P_{r}(k,t),
\end{equation}
where $N(k,t)$ is the number of nodes with degree $k$ at time $t$, and $P_{r}(k,t)$ is the
likelihood that a node immunized at time $t$ has a degree $k$. 

Then, plugging Eq.~\eqref{eq:pk of one removed node} into Eq.~\eqref{eq:N(t) first} gives, 
\begin{gather*}
N(k,t+1)=N(k,t)-\Delta \left[F(k,t)^{n}\right],
\end{gather*}
which becomes in the continuous limit,
\begin{gather*}
\frac{\partial N(k,t)}{\partial t}=-\Delta \left[F(k,t)^{n}\right].
\end{gather*}
Substituting $N(k,t)=(N-t)P(k,t)$, we get
\begin{gather*}
-P(k,t)+(N-t)\frac{\partial P(k,t)}{\partial t}=-\Delta \left[F(k,t)^{n}\right],
\end{gather*}
and using $P(k,t)=\Delta F(k,t)$, we obtain,
\begin{gather*}
\Delta \left[ -F(k,t)+(N-t)\frac{\partial F(k,t)}{\partial t} + F(k,t)^{n} \right]=0.
\end{gather*}
Note that $F(k=-1,t)=0$, and thus the entire term inside the $\Delta$ is $0$ for $k=-1$. Similarly, this implies that for $k=0$ and likewise for any $k\geq0$ this term is also $0$. Thus, we get the following simple ordinary differential equation,
\begin{equation}
(N-t)\dodo{}{t}F(k,t)=F(k,t)-F(k,t)^n,
\label{eq: ode}
\end{equation}
with the initial condition $F(k,t=0)=F(k)$. It can be shown (see Sec I in SM) that the solution of this ODE, Eq. (\ref{eq: ode}) is
\begin{equation}
F(k,t)=\Big(1+\big(F(k)^{1-n}-1\big)e^{(n-1)\log[(N-t)/N]}\Big)^{-\frac{1}{n-1}},
\end{equation}
or equivalently,
\begin{equation}
F_p(k)=\big(1+\left(F(k)^{1-n}-1\right)p^{n-1}\big)^{-\frac{1}{n-1}},
\label{eq:Fp(k)}
\end{equation}
where  $F_p(k)$ is the cumulative distribution of the degree after immunizing (removing) $1-p$ fraction of nodes. For $n=1$, the solution of Eq.~\eqref{eq: ode} is $F_p(k)=F(k)$ as expected. Also Eq.~\eqref{eq:Fp(k)} converges to $F(k)$ in the limit $n\to 1$. 

We can now obtain the degree distribution of the occupied nodes after fraction $1-p$ nodes are immunized, which is given by
\begin{equation}
P_p(k)=\Delta F_{p}(k)=F_{p}(k)-F_{p}(k-1).
\end{equation}
The equation for $v$, the probability of a randomly chosen link to lead to a node not in the giant component, is
\begin{equation}\label{equ:v1}
1-v=\sum_{k=0}^{\infty}\frac{kP(k)}{\av{k}}P({\Theta}|k)(1-v^{k-1}),
\end{equation}
where $P({\Theta}|k)$ is the probability of a node to be occupied given its degree is $k$. Using Bayes Theorem, we note that
 $P(k)P({\Theta}|k)=P({\Theta})P(k|{\Theta})=pP_p(k)$.
  Hence Eq. (\ref{equ:v1}) becomes
\begin{equation}\label{equ:v2}
1-v=\frac{p}{\av{k}}\sum_{k=0}^{\infty}kP_{p}(k)(1-v^{k-1}).
\end{equation}
The giant component $S$ can be found by
\begin{equation} \label{equ:Sgc}
S=\sum_{k=0}^{\infty}P(k)P({\Theta} |k)(1-v^{k})=p\sum_{k=0}^{\infty}P_p(k)(1-v^k),
\end{equation}
where $v$ is found from Eq. (\ref{equ:v2}).
 
At criticality, we take the derivative of both sides of Eq.~\eqref{equ:v2} and substitute $v=1$ representing the location where the first solution with $v<1$ exists, as opposed to only the $v=1$ solution. Thus, the critical condition is
\begin{equation}\label{equ:pc_simple}
1=\frac{p_c}{\av{k}}\sum_{k=0}^{\infty}k(k-1)P_{p_c}(k).
\end{equation}
\begin{figure}[h]
	\centering
	\includegraphics[width=4.2cm]{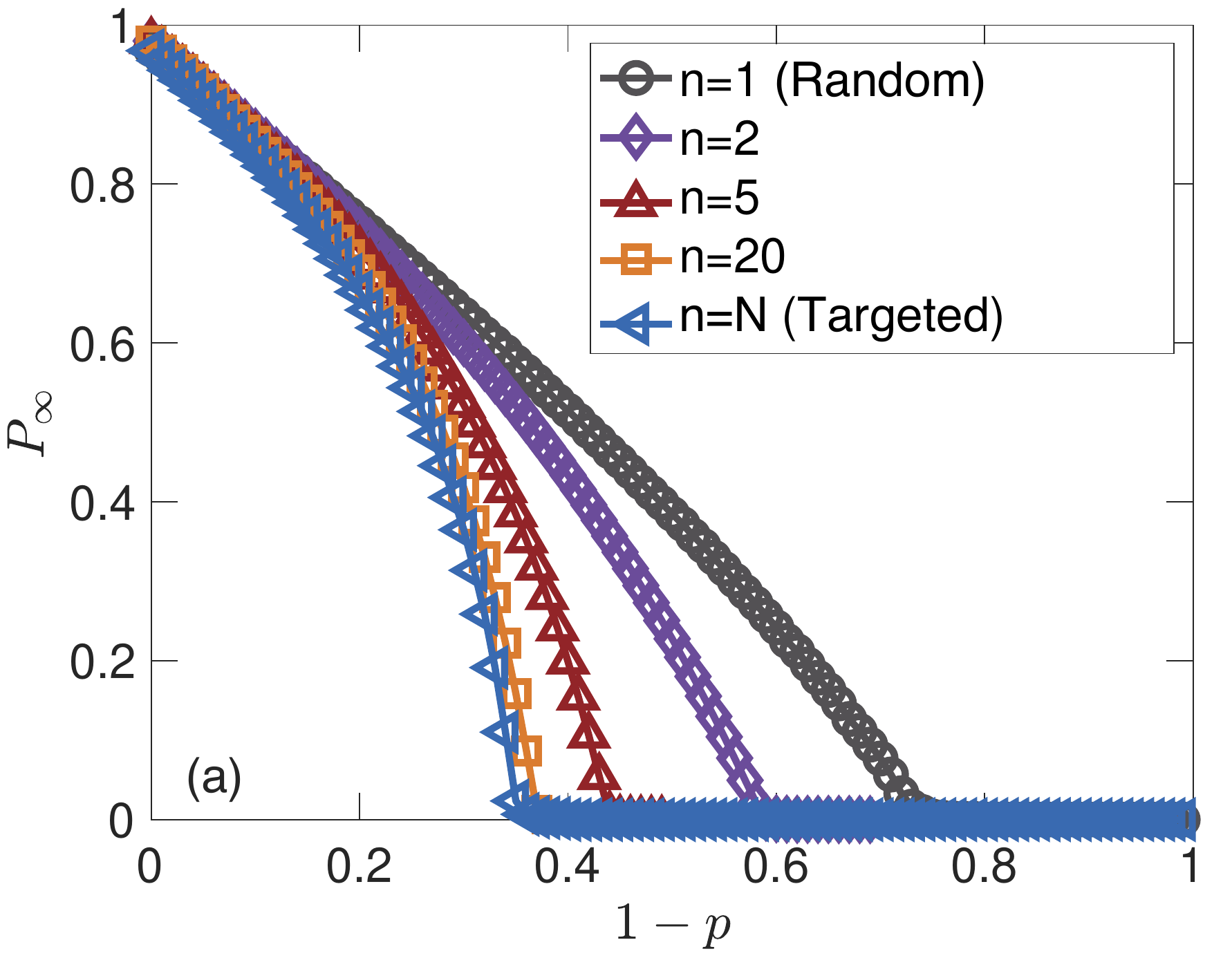}
	\includegraphics[width=4.2cm]{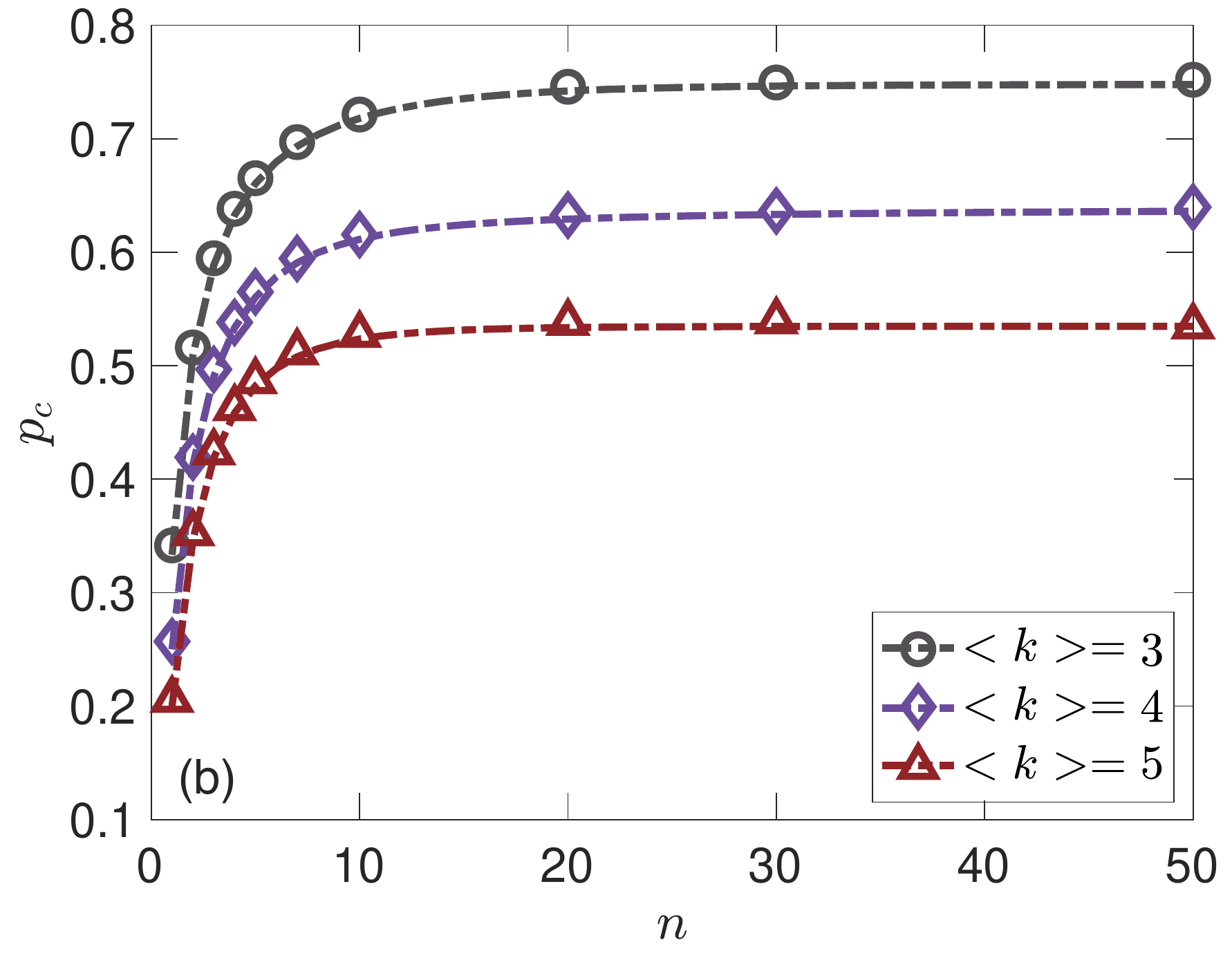}
	\includegraphics[width=4.2cm]{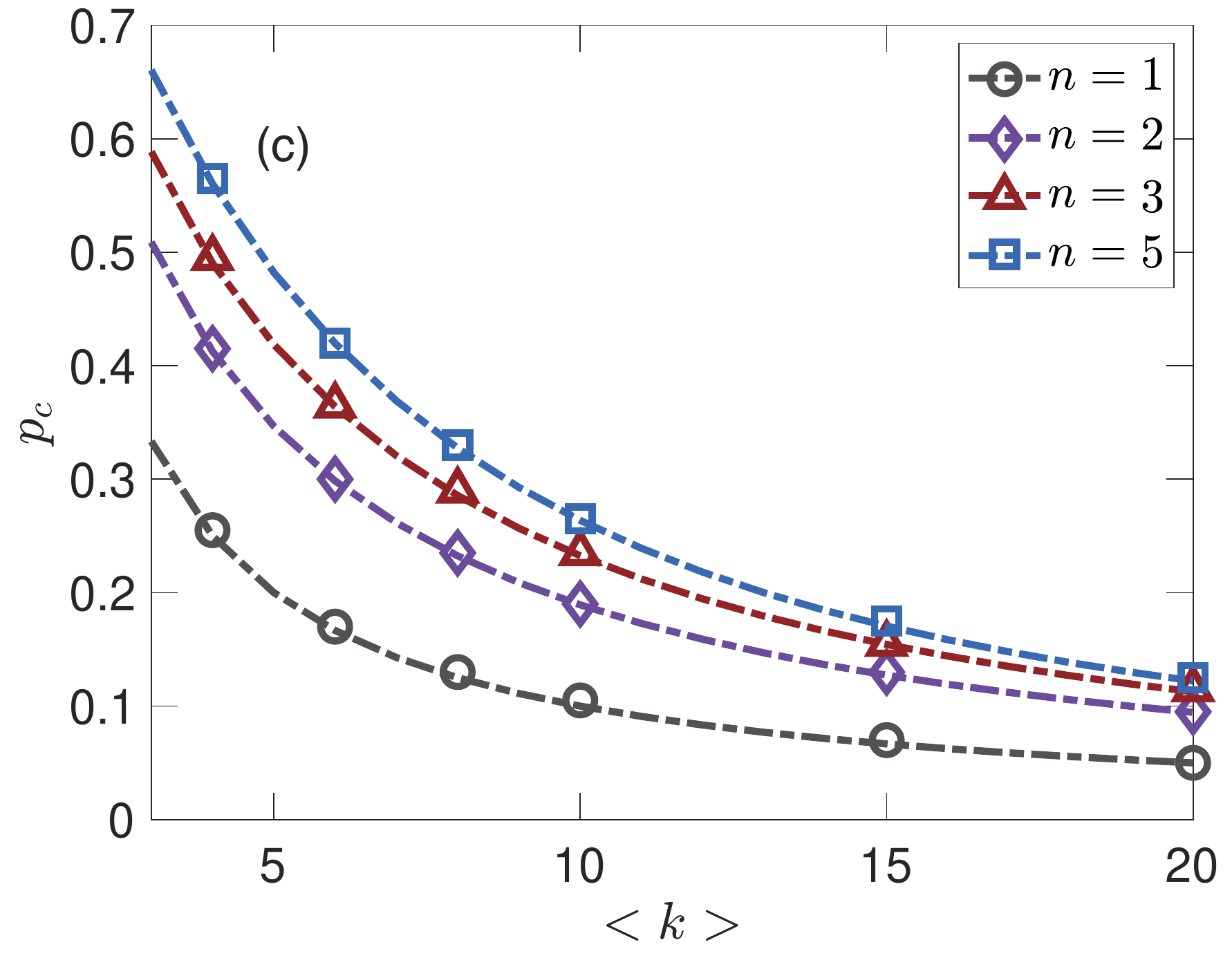}
	\includegraphics[width=4.2cm]{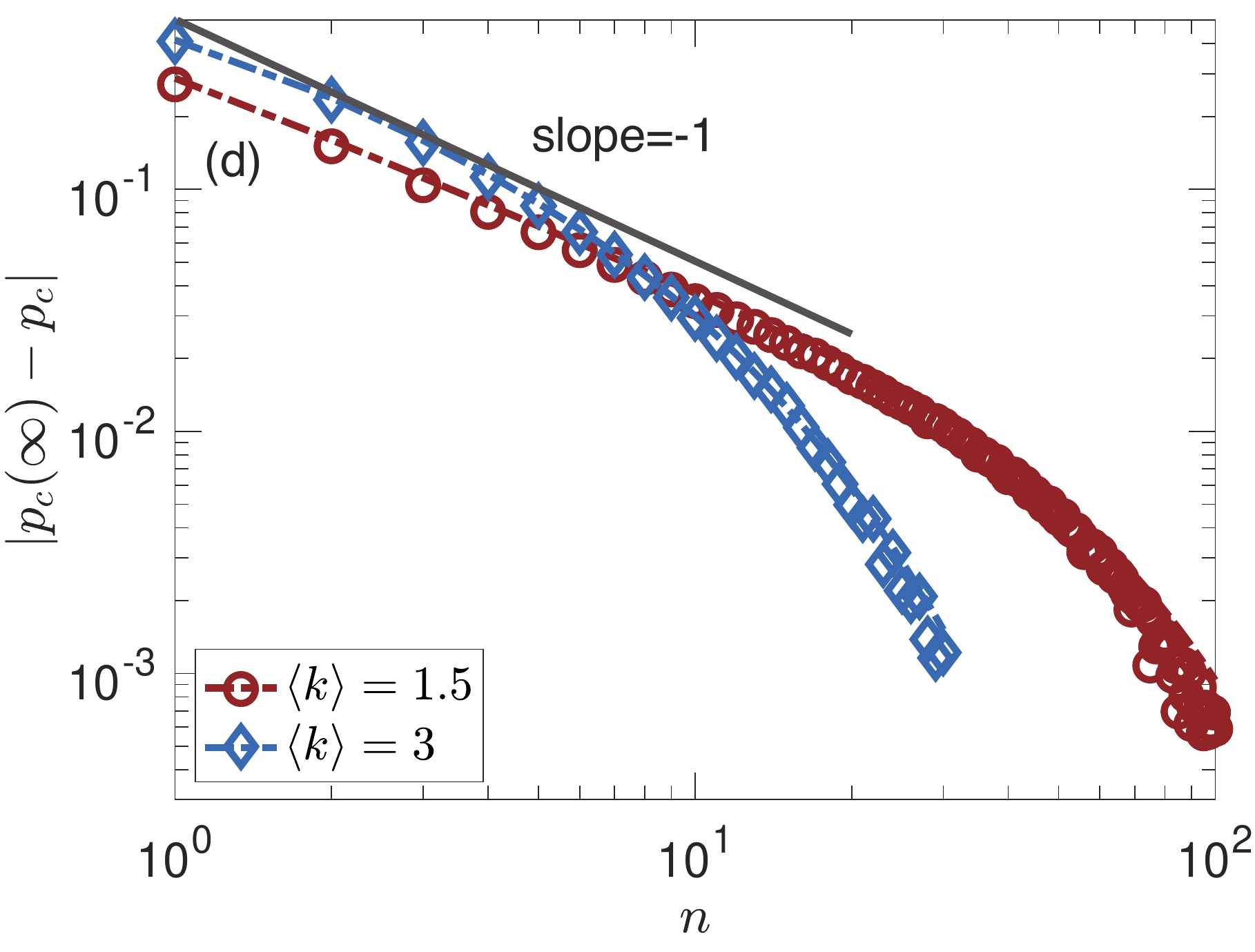}
	\caption{Results on ER networks. (a)
		The giant component $P_{\infty}$ of an ER network with $\av{k}=4$ varies with the fraction of immunized (or removed) nodes $1-p$ under limited knowledge. As the $n$ is increased, limited knowledge immunization tends to have the same immunization effect as targeted immunization. (b) The critical threshold $p_c$ of limited knowledge immunization as a function of $n$ on ER networks. Note that already for small $n\sim 10$, $p_c$ is close to targeted immunization (global knowledge, $n\sim N$). (c) Critical threshold $p_c$ as a function the mean degree $\av{k}$ of ER networks for limited knowledge immunization. (d) The scaling of $|p_{c}(\infty)-p_{c}|$ with $n$ on ER networks. Symbols are average results of simulations over $100$ independent realizations on ER networks with $10^6$ nodes. All simulation results (symbols) agree well with theoretical results from Eq.~\eqref{equ:pc_simple} (dashed lines). }\label{fig:ER}
\end{figure}

We now study our limited knowledge immunization (attack) strategy, i.e., the general result, Eqs. \eqref{equ:v2} and \eqref{equ:Sgc}, on ER networks. First, we analyze the giant component $P_{\infty}$. For the case $n=1$, limited knowledge immunization (or attack) reduces to the classical random attack, while for $n\to\infty$ (meaning the global network is observed) corresponds to targeted attack \cite{albert2000error,callaway2000network,cohen2001breakdown,huang2011robustness}. Using Eq. (\ref{equ:Sgc}), the giant component $P_{\infty}$ can be solved numerically for any given $p$. In Fig. \ref{fig:ER}(a), simulations and analytic results are shown for the giant component $P_{\infty}$ as a function of $1-p$ under limited information immunization with different $n$. As the knowledge index $n$ increases from $1$ to $N$, limited knowledge immunization moves from being like random immunization (attack) to being like targeted immunization (attack). The simulations are in excellent agreement with the theoretical results (lines).

Next, we focus on the critical threshold, $p_c$, of limited knowledge immunization (attack). Overall, we find that one does not need a very large knowledge $n\sim 10$ to achieve nearly the very close effect as targeted immunization (attack) with complete information. 
This can be seen by observing the critical threshold $p_{c}$ as a function of $n$ in Fig. \ref{fig:ER}(b). In Fig. \ref{fig:ER}(c) we show the variation of $p_c$ with $\av{k}$ for several fixed $n$. 

Next, the behavior of $p_c$ in the limit of large $n$ is derived analytically. 
By examining Eq.~\eqref{eq:Fp(k)} we notice that when $n\to \infty$ there are two distinct behaviors depending on whether $k$ is small, $F(k)<p$; or $k$ is large, $F(k)>p$. It can be shown (see Sec II in SM) that the leading term behaves as,
\begin{equation} \label{eq:Fp(k) large n}
F_p(k) \sim 
\begin{cases}
\frac{F(k)}{p}-\frac{1}{n} e^{-\alpha_kn}, & F(k)<p
\\
1-\frac{1}{n} e^{-\alpha_kn}, & p<F(k)<1
\\
1, & F(k)=1
\end{cases}
\end{equation}
where $\alpha_k=\left|\log \left[ p /F(k) \right] \right|$. In the limit $n\to\infty$, we can get the expected result for targeted immunization (attack), $F_p(k)=\min \left\{ F(k)/p,~1 \right\}$ \cite{callaway2000network,cohen2001breakdown}. 

Plugging Eq.~\eqref{eq:Fp(k) large n} into Eq.~\eqref{equ:pc_simple}, and noting that from a sum of exponentials decaying with $n$ only the lowest decay rate contributes to the leading term, we obtain (see Sec II in SM)
\begin{gather} \label{eq: pc vs n scaling}
p_c(n) \sim p_c^{\infty}  - A \frac{1}{n} e^{-\alpha n},
\end{gather}
where $p_c^{\infty}=p_c(n\to\infty)$, and the decay rate $\alpha$ is now
\begin{equation*}
	\alpha=\min_k \left|\log \left( p_c^{\infty} / F(k) \right) \right|.
\end{equation*}
The pre-factor $A=(2p_c^{\infty}k_{\rm slow})/(k_>  k_<)$, where 
$k_<$ is the largest degree such that $F(k)<p_c^{\infty}$,
$k_>=k_<+1$ and $k_{\rm slow}$ is the degree which gives the lowest rate $\alpha$. (See illustration in SM).

It is clear that $k_{\rm slow}$ must be $k_<$ or $k_>$ because $F(k)$ is monotonic. If $F(k_{\rm slow})=F(k_>)=1$ then $k_<$ should be taken as $k_{\rm slow}$, and the corresponding $\alpha$ should be taken. 
It should also be noted that if $k_{\rm slow}$ is not unique, it would simply change the pre-factor $A$ in Eq. \eqref{eq: pc vs n scaling}. Another special case is where $F(k_{\rm slow})=p_c^{\infty}$, then $|p_c^{\infty}-p_c| \sim 1 /n$ (see Sec IV in SM). 


Fig. \ref{fig:ER}(d) shows  $\Delta p_c=|p_c^{\infty}-p_{c}|$ as a function of $n$. As expected from the theory, one can see that $\Delta p_c\sim 1/n$ for small $n$ and exponential decay for large $n$. 
When $p_c\to 1$ which occurs for ER network when $\av{k}\to 1$, the power law regime becomes much broader as explained in the Sec II of SM. 

\begin{figure}
	\centering
	\includegraphics[width=4.2cm]{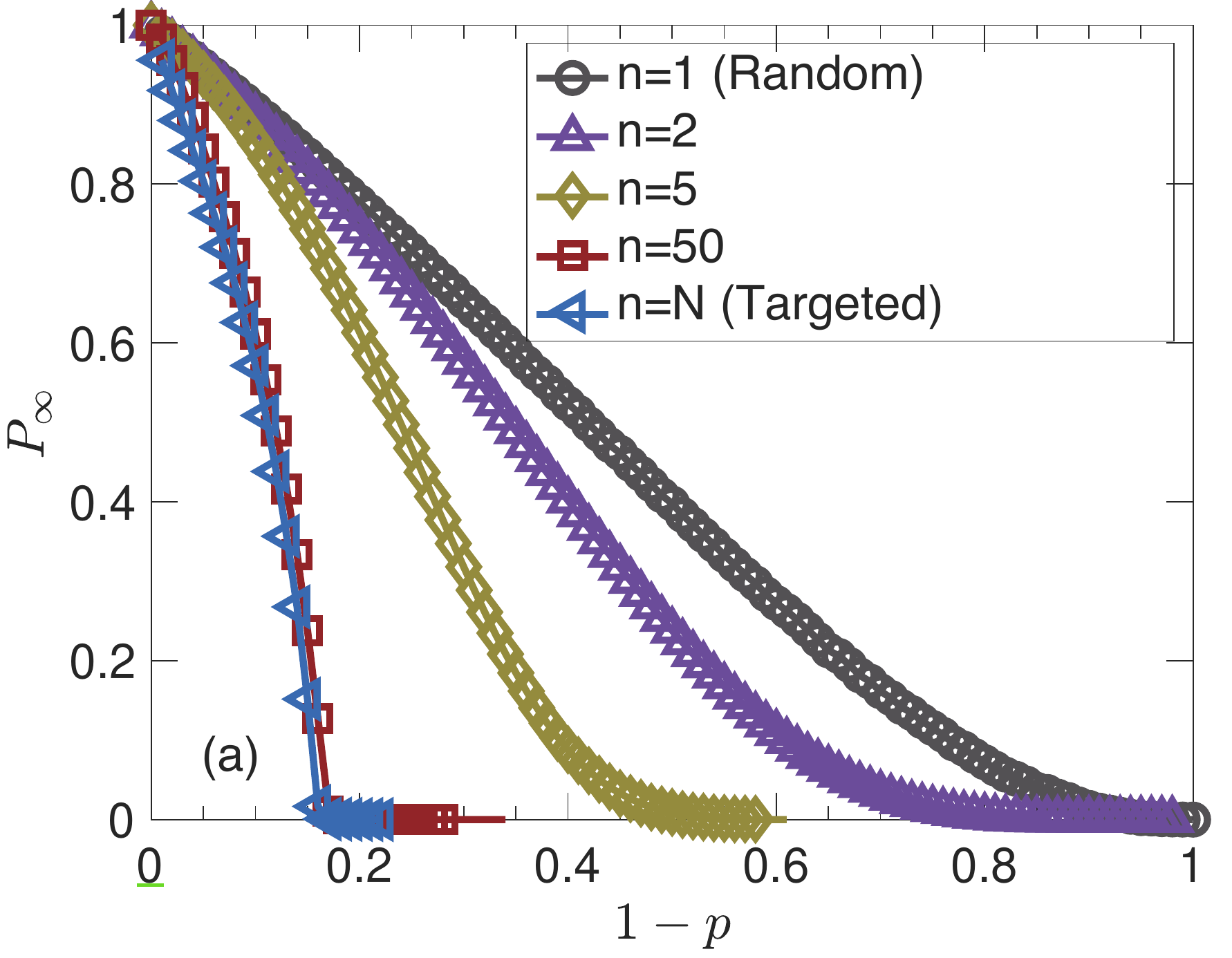}
	\includegraphics[width=4.2cm]{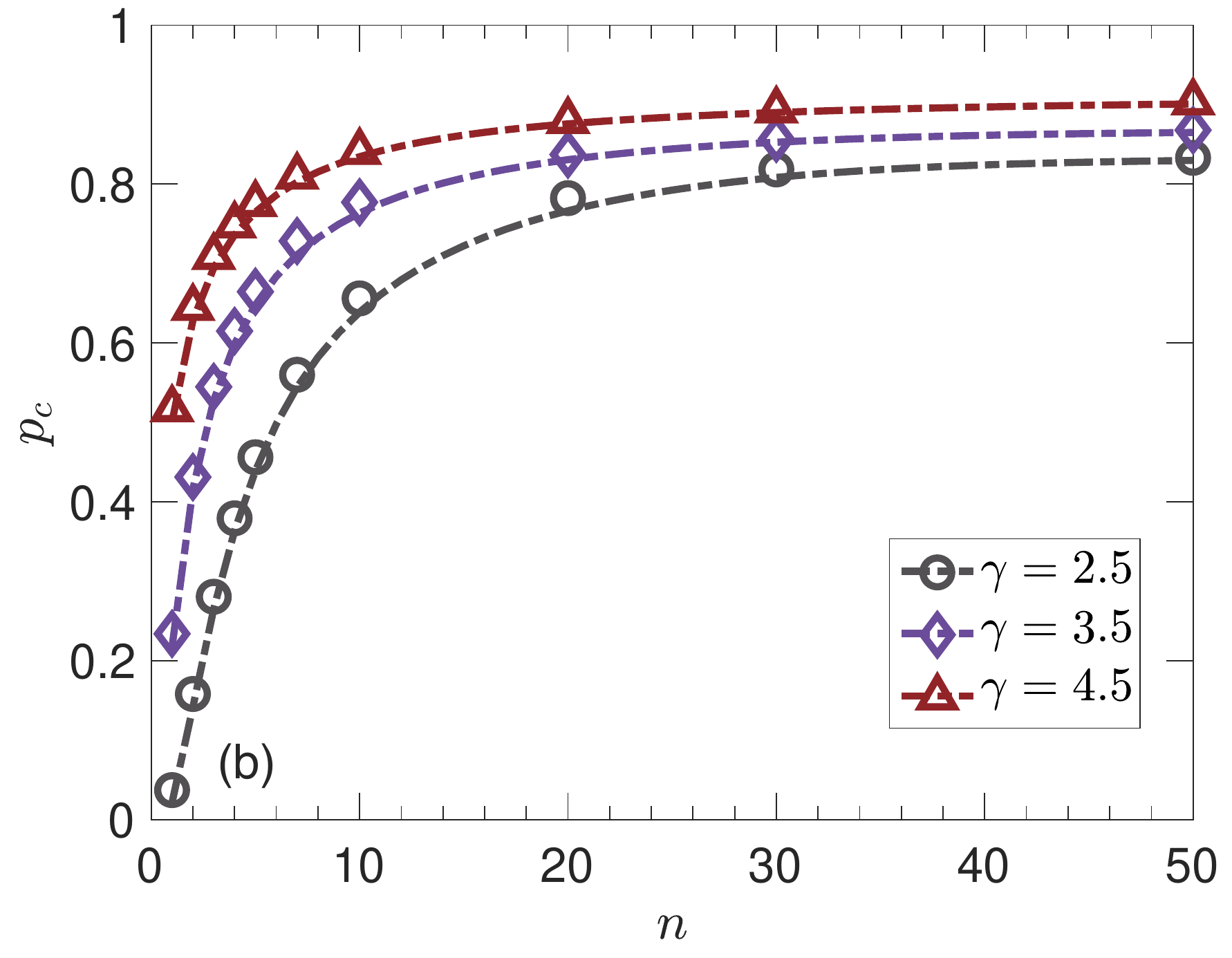}
	\caption{Results for SF networks. Comparison of theory (lines) and simulation (symbols) for limited knowledge immunization or attack, $n$, for SF networks. (a) The size of the giant component versus $n$ for SF network with $\gamma=2.5$. (b) Critical threshold $p_{c}$ versus $n$ on SF networks. Simulations are obtained for a system with $10^6$ nodes and averaged over $100$ independent realizations. The minimum and maximum degree of network are $m=2$ and $K=1000$ respectively.}\label{fig:SF}
\end{figure}

Next, we study SF networks with $P(k)=Ak^{-\gamma},k=m,\cdots, K$, where $A= (\gamma-1)m^{\gamma-1}$ is the normalization factor, and $m$ and $K$ are the minimum and maximum degree respectively \cite{cohen2000resilience}. Similar to ER networks, the size of the giant component, $P_\infty$ can be obtained from Eq. (\ref{equ:Sgc}). Fig. \ref{fig:SF}(a), shows $P_{\infty}$ as a function of $1-p$ for different $n$ values. The results demonstrate that  SF networks become more immunized/vulnerable compared to ER networks under the immunization/attack as $n$ increases. Compared with ER networks, one can observe that slightly higher values of $n$ (more knowledge) are needed to reach the near-steady-state region of fully targeted strategy. 

\begin{figure}
	\centering
	\includegraphics[width=0.9\linewidth]{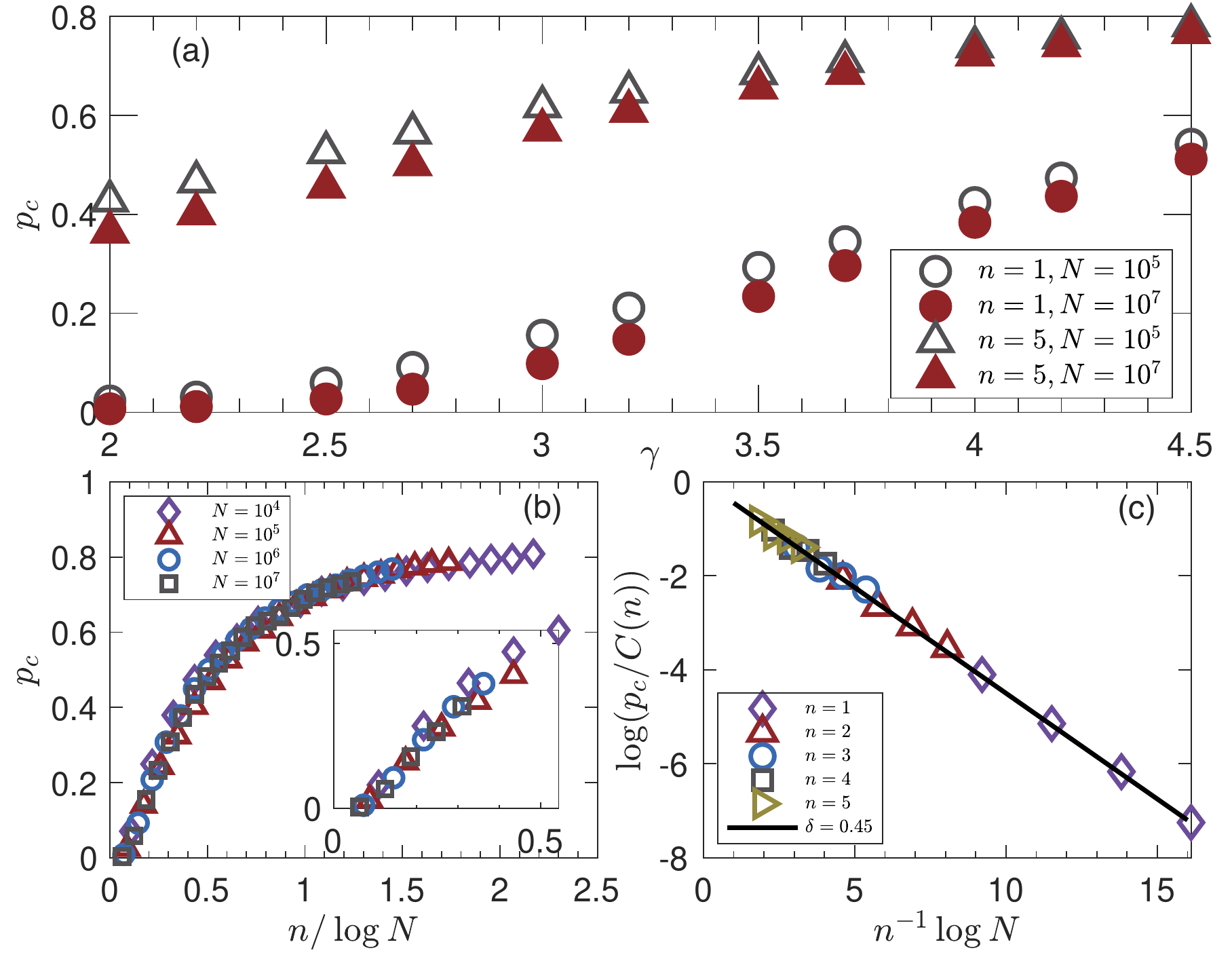}
	
	\caption{How $p_c$ for SF networks depends on $\gamma$ and the system size $N$. (a) $p_c$ as a function of $\gamma$ for different values of $n$ and $N$. $p_c$ decreases with increasing $N$. For $n=1$ and $2<\gamma<3$, it is well-known that $p_c$ approaches zero for infinite system. (b) $p_c$ as a function of $n/\log N$ for SF network with $\gamma=2.1$. The enlarged figure for small $n=1,\cdots, 5$ are shown in the inset, and $p_c$ approaches zero for $n/\log N \ll 1$. (c) The scaling of $p_c$ with $N$ and $n$ for large $N$ and small $n$. Here $C(n)$ is the pre-factor. The minimum and maximum degree of nodes are $m=2$ and $K=N^{1/(\gamma-1)}$ respectively. This confirms Eq.~\eqref{eq: pc vs N scaling SF} for $\gamma=2.1$, and $\delta=(3-\gamma)/2=0.45$.}
	\label{fig:4}
\end{figure}
For SF networks with $2<\gamma<3$, under random immunization/attack ($n=1$), it has been shown that $p_c=0$ for an infinite system \cite{cohen2000resilience}, while for high-degree immunization/attack ($n\to\infty$), $p_c>0$ \cite{callaway2000network,cohen2001breakdown}. Next we wish to find out for which $n$, $p_c$ becomes non-zero, and how it depends on the system size $N$. To this end, we analyze Eqs.~\eqref{eq:Fp(k)} and \eqref{equ:pc_simple} for large $k$ (high degrees govern the behavior in SF), small $n$ and $p$ as follows (elaborated in SM). 

It can be shown that for large degrees,
\begin{gather*}
F(k) \approx 1 - \left( k/m \right)^{1-\gamma}.
\end{gather*}
Substituting this into Eq. \eqref{eq:Fp(k)} and assuming $(k/m)^{\gamma-1} \gg n$ for large degrees, it can be concluded that
\begin{gather} \label{eq:Pp(k) SF}
P_p(k) \approx \dodo{F_p(k)}{k} \approx  p^{n-1} P(k).
\end{gather}
In addition, we notice that $P_p(k)$ has a new natural cutoff, $K_{p}$, which depends on $p$ and $N$ as follows (see Sec III in SM),
\begin{equation*}
	K_{p} \sim p^{n/(\gamma-1)} N^{1/(\gamma-1)}.
\end{equation*}
This helps us to evaluate the second moment of $P_p(k)$
\begin{gather*}
\av{k_{p}^2} \sim \int \limits_{m}^{K_{p}} k^2 p^{n-1} Ak^{-\gamma} \sim p^{n-1} {K_{p}}^{3-\gamma} \sim p^{n-1+n\beta} N^{\beta} ,
\end{gather*}
where $\beta=(3-\gamma)/(\gamma-1)$. 

Considering this, and
plugging Eq. \eqref{eq:Pp(k) SF} into Eq. \eqref{equ:pc_simple}, and keeping the leading terms in the limit of large $N$, we obtain (see Sec III in SM for more details)
\begin{gather} \label{eq: pc vs N scaling SF}
p_c \sim C(n) N^{-\delta/n} \sim C(n) \exp \left[ -\delta \frac{\log N}{n} \right],
\end{gather}
where
\begin{equation*}
\delta = \frac{\beta}{1+\beta} = \frac{3-\gamma}{2}.
\end{equation*}

From Eq. (\ref{eq: pc vs N scaling SF}), it is easy to see that if $n \ll \log N$ then $p_c \to 0$, while if $n\sim \log N$ then $p_c$ is non-zero. The pre-factor $C(n)$ depends on $n$ but not in $N$.

Fig. \ref{fig:4}(a) shows $p_c$ versus $\gamma$. It is known that for $2<\gamma<3$ and $n=1$, if $N\to \infty$ then $p_c \to 0$ \cite{cohen2000resilience}. Also for $n=5$, we can see that system size matters and $p_c$ decreases as $N$ increases.
Fig. \ref{fig:4}(b) shows that the scaling with $n/ \log N$ of Eq.~\eqref{eq: pc vs N scaling SF} is valid. Furthermore, it is seen in Fig.~\ref{fig:4}(b) that when $n$ is small or $N$ is large, such that $n/\log N \ll 1$ (in Fig.~\ref{fig:4} it is 0.07), $p_c$ approaches 0. Fig. \ref{fig:4}(c) supports the exponential scaling of $p_c$ versus $n^{-1}\log N$ obtained analytically in Eq. \eqref{eq: pc vs N scaling SF}.

In summary, our results provide a framework for understanding and carrying out efficient immunization with limited knowledge. Especially in cases of global pandemics such as e.g., the current COVID-19, it is impossible to know the full interactions of all individuals. Thus an effective way to limit spreading is obtaining information on a few ($n$) individuals and targeting the most central of these. For example, testers could stand at a supermarket and select a group of people entering the store simultaneously. Information on the connections of these people e.g., the number of people they live with, where and how often they meet with other people etc. could be quickly obtained (such as through cell phone tracking) and then the individual with the most connections in the group could be quarantined or immunized. Our results demonstrate that even when this is done in small groups of people (low $n$), it is possible to obtain a significant improvement in global immunization compared to randomly selecting individuals. In our model, this was seen by the reduced size of the giant component and the large critical threshold $p_c$. Overall, these findings could help to develop better ways for immunizing large networks and designing resilient infrastructure.


\begin{acknowledgments}

\end{acknowledgments}

\appendix

\bibliographystyle{unsrt}
\bibliography{mybib}

\end{document}